# PULSED DIFFERENTIAL SCANNING CALORIMETER MEASUREMENT OF COLLAGENE FIBER HEAT CAPACITY

M. Nadareishvili , T. Burjanadze. K. Kvavadze, N. Kekelidze, E. Kiziria , E.Gelagutashvili

*Iv. Javakhishvili Tbilisi State University*
*E. Andronikashvili Institute of Physics*
*0177, 6, Tamarashvili St.,*
*Tbilisi, Georgia*

Abstract

The work gives the estimation of the heat capacity jump at melting in water of collagen fibers of rat tail tendons. The measurements were carried out by an original calorimetric system designed by the authors, which in contrast to modern differential scanning calorimeters making the measurements under not equilibrium conditions due to the continuous heating of samples during the scanning, allows make the measurements both in scanning regime and also in pulsed regime under equilibrium conditions . The measurements carried out far from the phase transition in the pulsed regime allowed to determine exactly under equilibrium conditions the heat capacity before and after transition. As a result, the exact value of heat capacity jump under the equilibrium conditions was determined at melting.

## Introduction

The temperature of stability of collagen fibers exceeds the temperature of denaturation of triple helix in the solution by $22^0$C in average [1]. Though there is some success in understanding the nature of forces responsible for the increase of fibril stability, the problem is not studied completely yet. Some contribution to the increase of the temperature of fibril melting, as compared to triple helix, can be made by the hydrophobic interactions, which, as is known, are of entropy nature [2]. According to other explanation, the increase can be caused by additional hydrogen bonds between the triple helixes[3].
As it is known from the study of thermodynamic properties of globular proteins, for the solution of the role of hydrophobic interactions the most efficient approach is the measurement of heat capacity jump at the transition of protein from native to denaturation state [2]. Therefore, in this work the effort was made to estimate the value of heat capacity jump at melting the collagen fibers of rat tail tendons.
The existing differential calorimetric systems make the measurements only in scanning regime, as a result of which, the significant temperature gradients appear inside the measured cell and the pattern can be strongly deformed. The purpose of the paper is to study the thermodynamic properties of collagen fibers of at tail tendons under the conditions of thermodynamic equilibrium.

## Materials

**Rat tail tendon preparation**. Tails of six week-old rats were excised from the carcasses and released from skeleton. In order to exclude the side-effects connected with the change of the amount of hydrated water and with age-related changes [4], the experiments were made on fibrils from one and the same object. The fibrils were immediately washed in distilled water, and all visible contaminants were removed from fibrils under the microscope. Prior to experiments the fibrils were dried in vacuum for an hour at room temperature in order to remove the remained moisture. The collagen fibrils cleaned this way contain 97% of pure collagen. 10-20 mg fibril was placed into calorimetric ampoule and the distilled water was added, the amount of which was twenty times more as the weight of collagen, and was kept for a night before starting the experiment.

## Calorimetry

Measurements of a heat capacity jump of collagen fibers were carried out on a high-sensitivity and precision multipurpose pulsed differential scanning calorimeter (PDSC), which was designed in Andronikashvili institute of physics of Tbilisi Georgia, by authors. Calorimetric system PDSC differs from usual differential scanning calorimeters (DSC) that allows spend measurements not only in scanning mode, but also in a pulsed mode i.e. in equilibrium conditions.

At present the most sensitive calorimeters are the differential scanning calorimeters (DSC), they are characterized by comparatively shorter time of measurements also (that can be regulated by scanning rate), than the classical pulsed calorimeters and they are used more widely [5-8]. But they have one important disadvantage, since the measurement of heat capacities by these devices is made under non-equilibrium conditions because of continuous heating, and hence, they have not sufficient high accuracy (at high sensitivity), not allowing to study many "fine" effects. That is why, in case of DSC, the measured physical characteristics often depend on scanning rate [9].

Authors developed a new method of calorimetry to eliminate the main disadvantage of DSC (making measurements of the heat capacities under non-equilibrium conditions). On the basis of this method a pulsed differential calorimeter (PDC) of high-sensitivity and accuracy was created [10].

The measuring differential container of PDC consists of two identical cells, in which a sample and a standard are placed. The cells are connected by a thermal battery that measures the temperature difference between the cells and, at the same time, provides the required thermal link between them. Before applying a thermal impulse of $\Delta t$ duration and after passing a certain time of relaxation $\tau \gg \Delta t$, the studied sample and the standard are in thermal equilibrium and their temperatures are similar. During the whole process of measuring, the differential container is thermally isolated. The similar heat pulses $Q=IU\Delta t$ are applied to both cells by means of electric heater. The change of cell temperatures in time is described by the expressions:

$$T_1(t) - T_i = \frac{1}{C_1}\int_0^t \theta(\Delta t - t_1)\delta \dot{q}(t_1)dt_1 - \frac{k}{C_1}\int_0^t \delta T(t_1)dt_1$$

(1)

$$T_2(t) - T_i = \frac{1}{C_2}\int_0^t \theta(\Delta t - t_1)\delta \dot{q}(t_1)dt_1 + \frac{k}{C_2}\int_0^t \delta T(t_1)dt_1$$

where $T_i = T_1(0) = T_2(0)$ is the initial temperature of cells before applying the heat pulse, $\delta T(t_1)$ is the – instantaneous temperature difference between cell temperatures, k is the coefficient of heat conductivity between the cells, $\delta q(t_1)$ is the amount of heat power supplied to each cell during the action of the pulse and $\theta(t)$ is the step function determined by $\theta(t) = 1$ at $t \geq 0$ and $\theta(t) = 0$ at $t < 0$.

After finishing the relaxation at the moment $\Delta t + \tau$, exp. (1) becomes:
$$C_1 \Delta T = IU \Delta t - \Delta Q$$
$$C_2 \Delta T = IU \Delta t + \Delta Q, \qquad (2)$$

where $\Delta T = T_f - T_i$ is the temperature increment for the studied sample and the standard, $T_f = T_1(\Delta t + \tau) = T_2(\Delta t + \tau)$ is the final temperature of the cells after finishing the relaxation process, $IU\Delta t = \int_0^{\Delta t} \delta q(t) dt$ is the heat released in each cell during the action of the heat pulse, and $\Delta Q = k \int_0^{\Delta t + \tau} \delta T(t) dt$ is the amount of heat passed from one cell to the other (thanks to thermal link between the cells) due to temperature difference $\delta T(t)$ at the transition from one equilibrium state to other equilibrium state. And finally, the difference in heat capacities is calculated according to $\Delta C = 2\Delta Q/\Delta T$.

PDC has a very high sensitivity, as the heating rate in the pulse is high, but, at the same time, as compared to DSC, it has higher accuracy, as can make measurements under the equilibrium conditions like the classical pulsed calorimeters that measure the absolute heat capacity of samples. In the pulsed regime, PDC measures the difference in heat capacities $\Delta C$ between the studied and the reference samples under the equilibrium conditions, this difference can makes a very small (~1%) part of the heat capacity of the sample and is within the error of measurements of the calorimeters measuring the absolute heat capacities. The relative error $\delta C/C \approx 10^{-4}$, where $\delta C$ is the error of measurement of $\Delta C$, and C is the absolute heat capacity of the sample. Simultaneously with $\Delta C$, the absolute heat capacities of the studied and reference samples $C_1$ and $C_2$ respectively are measured with the same accuracy, as in the case of ordinary calorimeters measuring the absolute heat capacity.

By this calorimeter the high precision measurements of the anomaly of low-temperature heat capacity were made in different substances [10, 11].

PDC cannot be substituted for carrying out the experiments at low temperatures (T<50K), when the time $\tau = C/k$ of establishing the equilibrium in the studied system is short due to the low heat capacity of this system C and the large coefficient of heat conductivity k [12]. With the increase of temperature the time of pulse experiments sharply increases causing inconvenience. Therefore, the authors has combine pulse and scanning methods in one device and create a pulsed differential scanning calorimeter (PDSC), allowing to carry out the exact measurements with high sensitivity under the equilibrium conditions as in PDC, and at the same time requiring relatively short time for carrying out experiments, as in the case of DSC. It should be emphasized that this fact is very important. Such combination of regimes makes it possible to check the curves of heat capacity registered at relatively high temperature in the continuous heating regime, which, as our experience shows, frequently differ from the real curves. To determine the real temperature dependence of the heat effect at its measuring in continuous heating regime, it is enough to register the real values of this effect only at several temperatures in pulsed regime and thus, we can correct the whole curve. The sensitivity of the device in the operating temperature interval is ~$10^{-8}$ W, the working temperature interval of the device is 65K-370K, the rate of scanning at the continuous heating is possible to change from 6 K/h to 60 K/h. The loading of samples is made as usually without opening the vacuum system of calorimeter.

## Results and Discussion

The samples of collagen fibers were prepared the day before the experiment. In one of the ampoules 20 mg of fiber was placed and distilled water was added. In other ampoule the same amount of water was poured, the amount of water was controlled by analytical scales. Fig. 1 shows a typical pattern of temperature dependence of heat capacity at melting the rat tail tendon put into distilled water. In order to eliminate the additional errors connected with age-related changes and with other factors, all experiments were carried out on the fibrils of one and the same object. The results of measurements are given on the Fig. 2. Temperature of melting is T=331K, enthalpy, entropy and heat capacity jump at the denaturation in water are : $\Delta Hr=59.7 J/g$; $\Delta Sr=.179 J/gK$,. $\Delta C_p=0.46$. After denaturation, the heat capacity of collagen fibers does not depend on temperature within the experimental error and is equal 2.60 J/gK.

As it was mentioned above the value of heat capacity jump for fibrils in water makes 0.46J/gK. It may be caused by hydrophobic interactions, as in the literature there are data indicating the role of hydrophobic interactions at the formation of collagen fibers [13]. The relation between the value of heat capacity jump and the hydrophobic interactions is followed unambiguously on the example of globular proteins. For the majority of globular proteins, the hydrophobic interactions play the important role for the stability of globule. If we compare the results obtained by us with the results obtained for heat capacity jumps in globular proteins, in which the hydrophobic interactions really determine the stability of globule, its value in collagen is of the same order, as in ribonuclease - the protein with the lowest level of hydrophobic interactions [14].

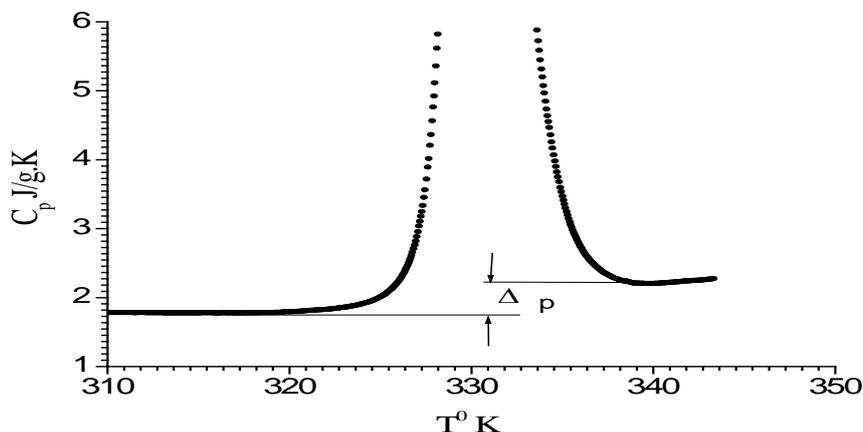

Fig.1 Typical calorimetric record of melting of collagen fibers in water

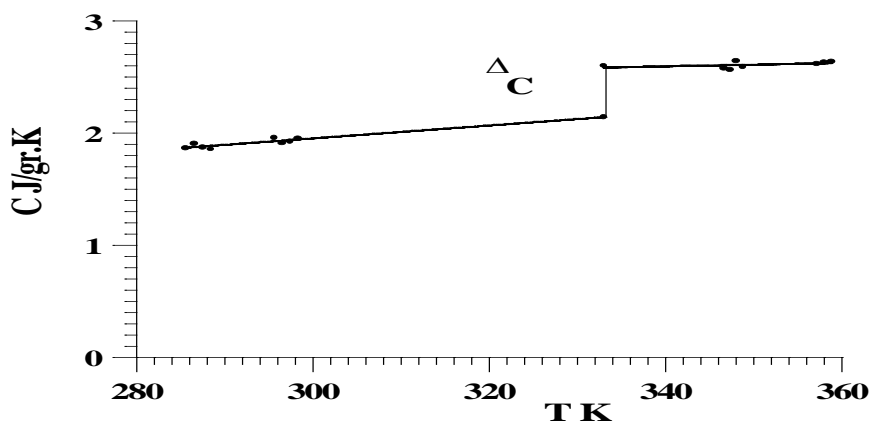

Fig.2. Partial heat capacity jump of collagen fibers at melting in water

**Summary**

The estimation of heat capacity jump at melting of collagen fibers having principle significance for determination of the role of hydrophobic interactions in the process of formation and stabilization of collagen fibers have been performed for collagen fibers of rat tail tendons in aqueous medium with high precision at thermodynamically equilibrium conditions what allows to except kinetic factors of melting a triple helix of collagen which can influence on this size. The researches have been done by a developed by authors high precision pulsed differential scanning calorimeter (Pulsed DSC) providing the measurements of heat capacity under the thermodinamically equilibrium conditions with high precision, in contrast to commonly used differential scanning calorimeters (DSC) making the measurements under not equilibrium conditions.

**References**

1. Rigby B.J. Thermal transitions in the Collagenous tissue of poikilotermic animals  J. Thermal biology 1977, **2**, 89-93.
2. Privalov P.L.  Stability of Proteins Edv. Prot.Chem, 1982, 35, 1-104.
3. Tiktopulo, E.I.  Kajava, A, V. Denaturation of Type I collagen fibers .Biochemistry 1998, 37, 8147-8152.
4. Miles, C.A. Ghelashvili, M. Polimer-in-a-Box mechanism for the thermal stabilization of collagen molecules in fiber. Biophysical Journal, 1999, **76**, 3243-3252.
5. P.L.Privalov, D.R.Monaselidze. Automatic adiabatic differential microcalorimeter for the investigation of structural transitions to macromolecules' .PTE, 1965. (Rus.)
6. G.P. Privalov, V. Kavina, E. Freire, P.L.Privalov "Precise scanning calorimeter for study of thermal properties of Biological macromolecules in dilute solution".Anal. Biochem., 1995, 232, 79-85.
7. Valerian V. Plotnikov, J. Michael  Brandts, Lung-Nan Lin, John F. Brandts. Anal.  Biochem., Academic Press, Ins. 1997, 250.
8. J. W. Loram, "A new high-preision continuous-heating differential calorimeter for the temperature range 1.5 K to 300 K.", J.Phys.E: Sci.Instrum., ,1983,.16, 367.   Printed   in Great Britain.
9. Miles, C.A., Burjanadze, T.V. Thermal stability of collagen fibers in ethylene glycol Biophysical Journal, 2001, 80**,** 1480-1486.
10. M.M. Nadareishvili, K.A. Kvavadze, G.G. Basilia, Sh.A. Dvali, Z.K horguashvili. „Excess Specific Heat of KCl Crystals Coused by Randomly distributed $OH^-$ Dipolar Impurities", Journal of Low Temperature Physics ., 2003, 130, Nos 5/6, 529-542.
11. G.G. Basilia, G.A. Kharadze, K.A. Kvavadze, M.M. Nadareishvili, D.F. Brewer, G.Ekosipedidis    and A.L. Thomson, Fizika Nizkih Temperatur, 1998, 24, No 8, 726-730
12. P.F. Sullivan and G. Seidel - Phys. Rev. 1968, 173, 679.
13. Schnell J. Evidences for the existence of hydrophobic interactions as a stabilizing factor in collagen structure. Arch.Bioch. Biophys. 1968, 127, 496-502.
14. Privalov P.L. Stability of  proteins. Small globular proteins. Edv. Protein Chem  1979, 33,167-241.